\documentclass[]{aastex}
\usepackage{emulateapj5,onecolfloat,graphicx}

\def\eg{{\it e.g.}}
\def\etal{{\it et al.}}
\def\etc{{\it etc.}}
\def\ie{{\it i.e.}}

\def\Ecirc{E_{\rm circ}}
\def\Eran{E_{\rm ran}}
\def\GCS{{\small GCS}}
\def\Hipp{{\small HIPPARCOS}}

\def\ILR{{\small ILR}}
\def\OLR{{\small OLR}}
\def\WASER{{\small WASER}}

\long\def\Ignore#1{\relax}

\def\pmb#1{\setbox0=\hbox{$#1$}%
  \kern-0.25em\copy0\kern-\wd0
  \kern.05em\copy0\kern-\wd0
  \kern-0.025em\raise.0433em\box0}

\slugcomment{Presentation on June 21, 2009 at ``Evolution of Planetary
  and Stellar Systems'' a conference in honor of Doug Lin, held in
  Prato, Italy}

\begin{document}

\twocolumn[
\title{New Developments in Spiral Structure Theory}
\author{J. A. Sellwood}
\affil{Rutgers University, Department of Physics \& Astronomy, \\
       136 Frelinghuysen Road, Piscataway, NJ 08854-8019 \\
       {\it sellwood@physics.rutgers.edu}}

\begin{abstract}
After a short review of the principal theories of spiral structure in
galaxies, I describe two new developments.  First, it now seems clear
that linear theory cannot yield a full description for the development
of spiral patterns because $N$-body simulations suggest that non-linear
effects manifest themselves at a relative overdensity of $\sim 2\%$,
which is well below the believed spiral amplitudes in galaxies.
Second, I summarize the evidence that some stars in the solar
neighborhood have been scattered at an inner Lindblad resonance.  This
evidence strongly supports a picture of spirals as a recurring cycle
of transient instabilities, each caused by resonant scattering by a
previous wave.
\end{abstract}

]

\section{Introduction}
Despite many decades of effort, we do not yet have a widely accepted
theory to account for the graceful spiral patterns in disk galaxies.
Most workers in this field agree that spiral patterns are
gravitationally driven variations in surface density in the old
stellar disk, as supported by photometric data \citep[\eg][]{ZCR9} and
streaming motions in high spatial resolution velocity maps
\citep[\eg][]{SVOT}.  There is also general agreement that gas seems
to be essential.

There seems little doubt that some spiral patterns are tidally driven,
while others could be the driven responses to bars.  Although these
two ideas may account for a large fraction of the cases, especially if
orbiting dark matter clumps can excite patterns \citep{Dubi08},
spirals can still develop in the absence of either trigger, as
revealed in simulations.

Here, I focus on the idea that spirals are self-excited oscillations
of the stellar disk, which represents the greatest theoretical
challenge.  Two camps have long advanced quite distinct theories.  One
idea \citep[\eg][]{BL96}, is that spiral features are manifestations
of quasi-steady global modes of the underlying disk.  The other is
that they are short-lived, recurrent transient patterns that originate
either from shearing bits and pieces \citep[\eg][]{Toom90}, or
something more subtle \citep[\eg][]{Sell00}.

A serious barrier to progress in this area has been the absence of
observational discriminants that would favor one of these radically
differing viewpoints over the other.  The predictions for density
variations or gas responses at a single instant are essentially
independent of the generating mechanism and do not depend strongly on
the lifetime of the pattern.

In his presentation, but not in the written version, \cite{Toom81}
likened theoretical work on spiral structure to the (apocryphal) blind
men examining an elephant.  Several important features of the beast
have been described in papers by many different authors over the past
50 years.  Here I first review many of these components before
attempting to weave them into a distinct, yet still incomplete,
picture.

The key idea described in this review resulted from a collaborative
visit to Doug Lin in Santa Cruz in the summer of 1987.  I developed it
further for a few years, but worked on other problems whilst awaiting
observational support for the somewhat complicated picture we
proposed.  The desired evidence appeared with the publication of the
Geneva-Copenhagen Survey \citep[][hereafter \GCS]{Nord04}, as I will
explain below, and results from the {\it Gaia} satellite should
improve matters still further.

\section{Global modes of rotationally-supported disks}
Simple models of disk galaxies support many linear instabilities
\citep[\eg][]{Korc05,Jala07}.  The bar-forming mode is generally the
fastest growing, but it has almost no spirality.  These studies are
therefore important to understand stability, but do not appear
promising for spiral generation.

The ``density wave'' theory for spiral modes, described in detail by
\cite{BL96}, invokes a more specific galaxy model with a cool outer
disk and hot inner disk.  The local stability parameter, $Q = \sigma_R
/ \sigma_{R,\rm crit}$ \citep{Toom64}, is postulated to be $Q \ga 1$
in the outer disk and to rise steeply to $Q > 2$ near the center.
Under these specific conditions, Bertin \& Lin find slowly evolving
spiral modes that grow by their \WASER\ mechanism.  They invoke shocks
in the gas to limit the amplitude of the slowly growing mode, leading
to a quasi-steady global spiral pattern.  \cite{Lowe94} present a
model of this kind to account for the spiral structure of M81.  The
main objection to their picture is that it is likely that such a
lively outer disk will support other, more vigorous, collective
responses that will quickly alter the background state by heating the
outer disk, as described below.

\section{Recurrent transients}
$N$-body simulations of cool, shearing disks always exhibit recurrent
transient spiral activity, and this basic result has not changed for
several decades as numerical quality has improved.  \cite{SC84}
reported that patterns fade in simple simulations that do not include
the effects of gas dissipation; the reason is the disk becomes less
responsive as random motion rises due to particle scattering by the
spiral activity \citep{CS85,BL88}, which is therefore self-limiting.
\cite{SC84} also show that mimicking the effects of dissipative
infall of gas, such as by adding fresh particles on circular orbits,
allowed patterns to recur ``indefinitely.''  Later work
\citep[\eg][]{CF85,Toom90,Chak08} has shown that almost any method of
dynamical cooling can maintain spiral activity, as also happens in
galaxy formation simulations \citep[\eg][]{Gove07}.

\section{Indirect evidence for Transient Spirals}
Thus the transient spiral picture offers a natural explanation for the
absence of spiral patterns in S0 disk galaxies that have little or no
gas; maintenance of spiral activity requires a constant supply of new
stars on near-circular orbits.  There are several other indirect
pieces of evidence that also favor the transient spiral picture.

It has been clear for sometime that the velocity dispersion of disk
stars in the solar neighborhood rises with age (\eg\ \GCS) and also,
for main sequence stars, with color \citep{AB09}, which is a surrogate
for mean age.  The highest velocities cannot be produced by cloud
scattering \citep{Lace91,HF02} and some other accelerating agent, such
as transient spirals, seems to be required \citep{BL88}.

Studies of the metallicities and ages of nearby stars
\citep{Edva93,Nord04,Reid07,Holm07} find that older stars tend to have
lower metallicities on average.  As it is difficult to estimate the
ages of individual stars, the precise form of the relation is still
disputed.  However, there seems to be general agreement that there is
a spread of metallicities at each age, which is also supported by other
studies \citep{CHW03,Hayw08}.  A metallicity spread amongst coeval
stars is inconsistent with a simple chemical evolution model in which
the metallicity of the disk rises monotonically in each annular bin,
without mixing in the radial direction \citep{SB09}.  Again
\cite{SB02} showed that the needed radial mixing arises naturally
when the disk supports recurrent transient spirals.  (See also Ro\u
skar \etal\ 2008ab).

\begin{figure}[t]
\includegraphics[width=.9\hsize]{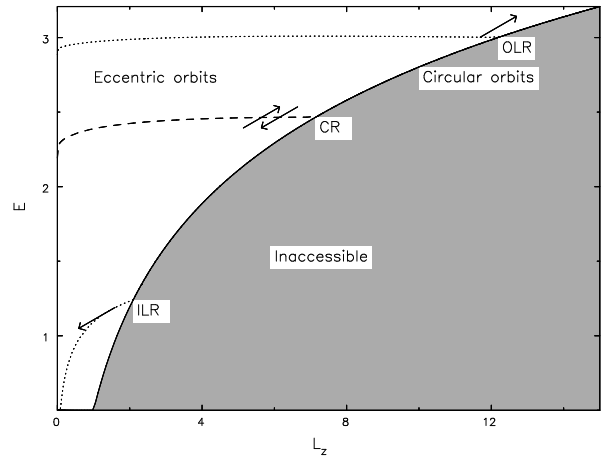}
\caption{The Lindblad diagram for a disk galaxy model.  Circular
  orbits lie along the full-drawn curve and eccentric orbits fill the
  region above it.  Angular momentum and energy exchanges between a
  steadily rotating disturbance and particles move them along lines of
  slope $\Omega_p$ as shown.  The dotted and dashed lines show the
  loci of resonances for an $m=2$ perturbation of arbitrary pattern
  speed.}
\label{lindblad}
\end{figure}

\section{Wave-Particle Scattering}
Figure~\ref{lindblad} shows, for an axisymmetric disk having a flat
rotation curve, how the classical integrals of specific energy, $E$,
and angular momentum, $L_z$, are changed for stars that are scattered
by a steadily rotating mild potential perturbation.  The solid curve
shows the locus of circular orbits, which has slope $\Omega_c$, while
stars with more energy for their angular momentum pursue non-circular
orbits.  The principal resonances for an $m$-fold rotationally
symmetric disturbance that rotates at angular rate $\Omega_p$ are
marked.  The familiar corotation and Lindblad resonances for near
circular orbits can be generalized for non-circular orbits by using
the angular frequencies, $\Omega_\phi$ \& $\Omega_R$, of their doubly
periodic motion defined in \cite[][hereafter BT08, pp~146-147]{BT08}.
For nearly circular orbits, $\Omega_\phi \rightarrow \Omega_c$ and
$\Omega_R \rightarrow \kappa$, which are the angular frequencies of
circular and epicycle motion respectively.  The dotted curves show the
loci of the principal resonances where $m(\Omega_\phi - \Omega_p) =
l\Omega_R$; here $l=0$ for corotation and $l=\pm1$ of the inner and
outer Lindblad resonances respectively.

\cite{LBK2} show that stars are scattered by spiral waves only at
resonances.  Since Jacobi's constant (BT08, eq.~3.112), is conserved
in axes that rotate with the perturbation, the slope of all scattering
vectors in this plot is $\Delta E / \Delta L_z = \Omega_p$.  Since
$\Omega_c = \Omega_p$ at corotation, scattering vectors are parallel
to the circular orbit curve at this resonance, where angular momentum
changes do not alter the energy of random motion.  Outward transfer of
angular momentum involving exchanges at the Lindblad resonances, on
the other hand, does extract energy from the potential well that is
converted to increased random motion, as is well known.

\cite{SB02} show that scattering at corotation causes very effective
mixing.  In a few Gyr, multiple transient spirals cause stars to
diffuse in radius over time.  These changes at corotation of the
spirals, which occur with no associated heating, are able to account
for the apparent metallicity spread with age.

Notice also from Fig.~\ref{lindblad} that the direction of the
scattering vector closely follows the resonant locus (dotted curve) at
the \ILR\ only.  Thus, when stars are scattered at this resonance, they
stay on resonance as they gain random energy, allowing very strong
scatterings to occur.  The opposite case arises at the outer Lindblad
resonance (\OLR), where the star quickly moves off resonance when it
gains a small amount of angular momentum.

\section{Swing Amplification}
The classic figure of dust to ashes presented by \cite{Toom81} and
reproduced in BT08 (Fig.~6.19) shows the dramatic transient trailing
spiral that results from a small input leading disturbance.  This
linear calculation shows that the disturbance does not persist, but
decays after its transient flourish.  In the late stages, the ``wave
action'' propagates at the group velocity \citep[][and BT08
pp~499-500]{Toom69} away from corotation, where it is absorbed at the
Lindblad resonances.

\cite{Toom90} suggests that a large part of the spiral activity
observed in disk galaxies is the collective response of the disk to
shot noise in the density distribution.  \cite{TK91} show that
amplified noise arising from the massive disk particles themselves can
be understood in the shearing sheet, where the resulting spiral
amplitudes are linearly proportional to the input level of shot noise.
While this may be a mechanism for chaotic spirals in very gas rich
disks, where the clumpiness of the gas distribution may make the seed
noise amplitude unusually high, it seems likely that spiral amplitudes
\citep[\eg][]{ZCR9} are too large to be accounted for by this
mechanism in most galaxies.  Also, the spiral structure should be
chaotic, with little in the way of clear symmetry expected.

\section{Groove modes}
\cite{SK91} showed that disks are destabilized by a deficiency of
stars over a narrow range of angular momentum, which creates a
``groove'' in a disk without random motion.  The groove itself is
unstable, and the instability becomes a global mode through the
vigorous supporting response of the surrounding stellar disk.  The
resulting linear instability, which also develops in a disk with
random motion, produces a large-scale spiral pattern.  \cite{SB02}
showed that the amplitude of the mode was limited by the onset of
horseshoe orbits at corotation.

\cite{SK91} showed that almost any narrow feature in the angular
momentum density is destabilizing.  Thus the common starting
assumption of spiral structure studies, that the underlying disk is
featureless and smooth, may throw the spiral baby out with the
bathwater.

\begin{figure}[t]
\includegraphics[width=.9\hsize]{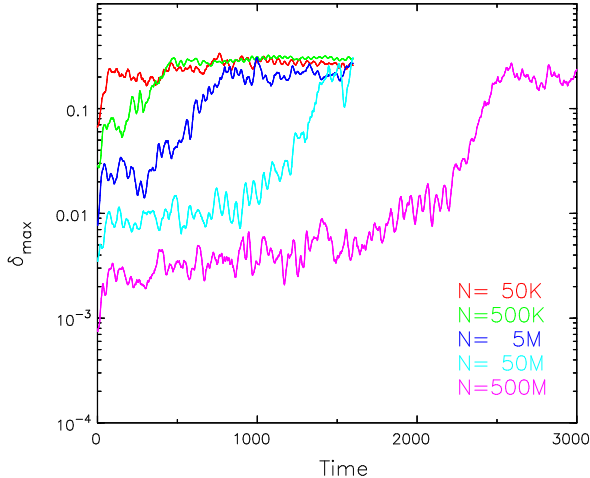}
\caption{The time evolution of the peak overdensity in a series of
  simulations of the half-mass Mestel disk with different numbers of
  particles.  The model is predicted by \cite{Toom81} to be globally
  stable.  The ordinate is the maximum values of $\delta$ on grid
  rings.  Linear theory predicts the amplitude should remain
  proportional to $N^{-1/2}$.}
\label{cntrst}
\end{figure}

\section{A linearly stable disk}
The so-called ``Mestel'' disk has the scale-free surface density
profile $\Sigma = V_c^2/(2\pi GR)$, with $V_c$ being the circular
orbital speed that is independent of $R$.  \cite{Zang76} and
\cite{Toom81} carried through a global, linear stability analysis of
this disk model with random motions described by a smooth distribution
function.  They introduced a central cutout and an outer taper in the
active mass density, and replaced the removed mass by rigid components
in order that the central attraction remained unchanged at all radii.
The dominant linear instabilities they derived were confirmed in
$N$-body simulations by \cite{SE01}.

\cite{Zang76} and \cite{ER98} showed that all full-mass disk models
suffered from global lop-sided instabilities no matter how large a
degree of random motion or however gentle the cutouts.  Accordingly,
\cite{Toom81} halved the active disk mass and was able to prove that
the resulting model was globally stable to all small-amplitude
disturbances when $Q=1.5$.  This is the only known model for a
non-uniformly rotating disk that is stable to small amplitude
perturbations.

Linear theory therefore predicts that $N$-body simulations of the
half-mass Mestel disk should exhibit no activity in excess of the
inevitable swing-amplified shot noise.  Figure~\ref{cntrst} shows that
this is not the case.  The ordinate is the largest value of $\delta =
\Delta \Sigma/ \Sigma$ from $m = 2$ disturbances in a sequence of
simulations with increasing numbers of particles.  The unit of time is
$R_i/V_c$, where $V_cR_i$ is the center of the inner angular momentum
cut out.  Thus the orbit period at this small radius is $2\pi$.

At $t=0$, $\delta \propto N^{-1/2}$, as appropriate for shot noise,
and swing-amplification causes an almost immediate jump by a factor
$\la 10$ for all $N$.  When $N = 50$K, amplified noise causes $\sim
20\%$ overdensities, but the amplitude always rises to similar values
in later evolution, when the inner disk has developed a pronounced
bar, for all the particle numbers shown.  A period of slow growth
occurs in the simulations with the largest two values of $N$, offering
tentative support for the linear theory prediction of global
stability, with the slow growth perhaps arising from the gradual
growth of particle correlations as described by \cite{TK91}.  Even in
these cases, however, the amplitude rises more rapidly once $\delta
\ga 2\%$.

\begin{figure}[t]
\includegraphics[width=.9\hsize]{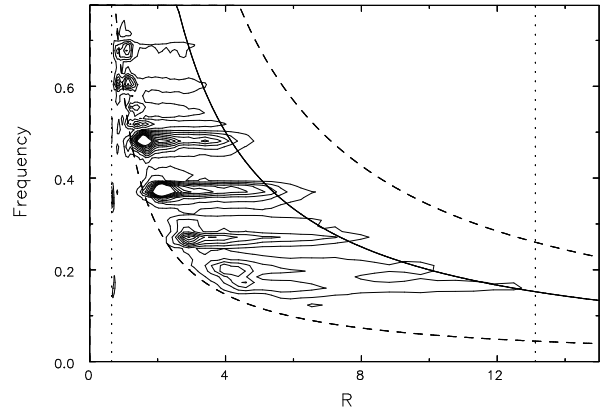}
\caption{A power spectrum of $m=2$ disturbances during the period of
  rapid rise in amplitude in the $N=5 \times 10^7$ experiment from
  Fig.~\ref{cntrst}.  The solid curve shows $2\Omega(R)$ and the
  dashed curves $2\Omega(R) \pm \kappa(R)$.}
\label{spctrm}
\end{figure}

Thus we always observe runaway growth of spiral activity -- albeit
more and more delayed as $N$ is increased -- behavior that is quite
clearly not in accord with linear theory predictions.  The rapid
growth once $\delta \ga 2\%$ suggests that the behavior is already
non-linear at this modest amplitude.  Note that the largest number of
particles, $N = 500$M, is within a factor of 100 of the number of
stars in a real galaxy disk, where in reality the mass distribution is
far less smooth, owing to the existence of star clusters and giant gas
clouds.

\section{More detailed analysis}
The runaway growth once $\delta \ga 2\%$ is of particular interest.
A power spectrum analysis of non-axisymmetric density variations during
this period of rapid growth reveals multiple waves of many different
frequencies, each extending from the inner Lindblad resonance (\ILR) to
corotation, as shown in Fig.~\ref{spctrm}.  We see that disturbances
are present at several frequencies at any one time, and that each
appears and fades over a period of $\sim 100$ dynamical times, with
those of lower frequency generally developing later.

\begin{figure}[t]
\includegraphics[width=.9\hsize]{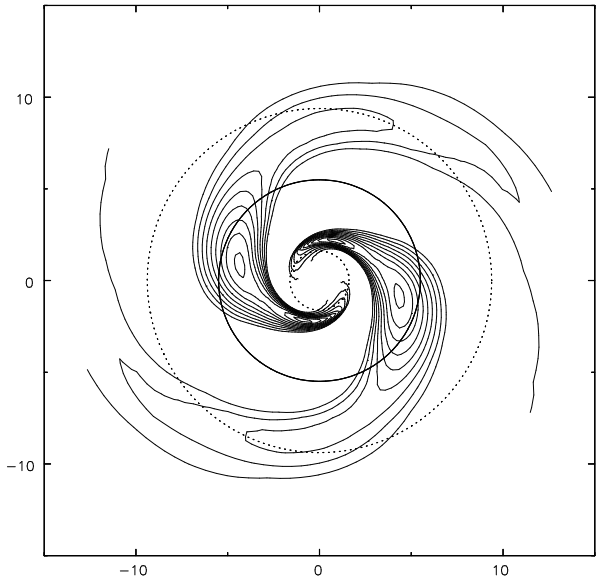}
\caption{The optimal coherent wave fitted to the $m=2$ data from the
  simulation described in \S 1.5.  The full-drawn circle marks the
  corotation resonance, the dotted circles the Lindblad resonances for
  this pattern.}
\label{mode}
\end{figure}

Figure~\ref{mode} shows the transient wave that was prominent over a
short time interval during which the wave grew and decayed while the
pattern speed remained approximately constant.  The measured $\Omega_p
= 0.182V_0/R_i$, placing corotation and the Lindblad resonances for
circular orbits at the radii shown by the circles in this Figure.  The
peak amplitude of the pattern is near corotation and the wave extends
to the Lindblad resonances on either side; the outer half has lower
amplitude because the wave is spread over a larger area.  As the wave
decays, the corotation peak disperses \citep{SB02} and the ``wave
action'' is carried radially at the group velocity where it is
absorbed at the Lindblad resonances.

It is convenient to separate the specific energy of a star in an
axisymmetric potential, $\Phi(R,z)$, into three distinct terms $
E = \Ecirc(L_z) + \Eran + E_z$, where
\begin{eqnarray}\nonumber
\Ecirc & = & \Phi(R_c,0) + L_z^2/(2R_c^2) \cr
\Eran & = & \Phi(R,0) - \Phi(R_c,0) + {1\over2}\left[v_R^2 + L_z^2\left({1\over R^2} - {1\over R_c^2}\right)\right] \cr
E_z & \simeq & {\textstyle{1\over2}}(z^2\nu^2 + v_z^2). \cr
\label{deferan}
\end{eqnarray}
Here $R_c$ is the radius of a circular orbit having the same $L_z$, and
$v_R$ is the radial velocity.  The approximate form for $E_z$ (BT08,
eq.~3.86) holds for stars that do not climb to great heights above the
mid-plane, with $\nu$ being the vertical oscillation frequency of such
stars.

The change in random energy caused by an angular momentum change
$\Delta L_z$ is $\Delta \Eran = \Omega_p \Delta L_z - \Ecirc(L_z +
\Delta L_z) + \Ecirc(L_z)$.  It is easy to show that scattering from a
circular orbit according to this formula takes place along a line in
$(L_z,\Eran)$-space that is has a steep negative slope for the \ILR\
and positive slope for the \OLR.

\begin{figure}[t]
\includegraphics[width=.9\hsize,clip=]{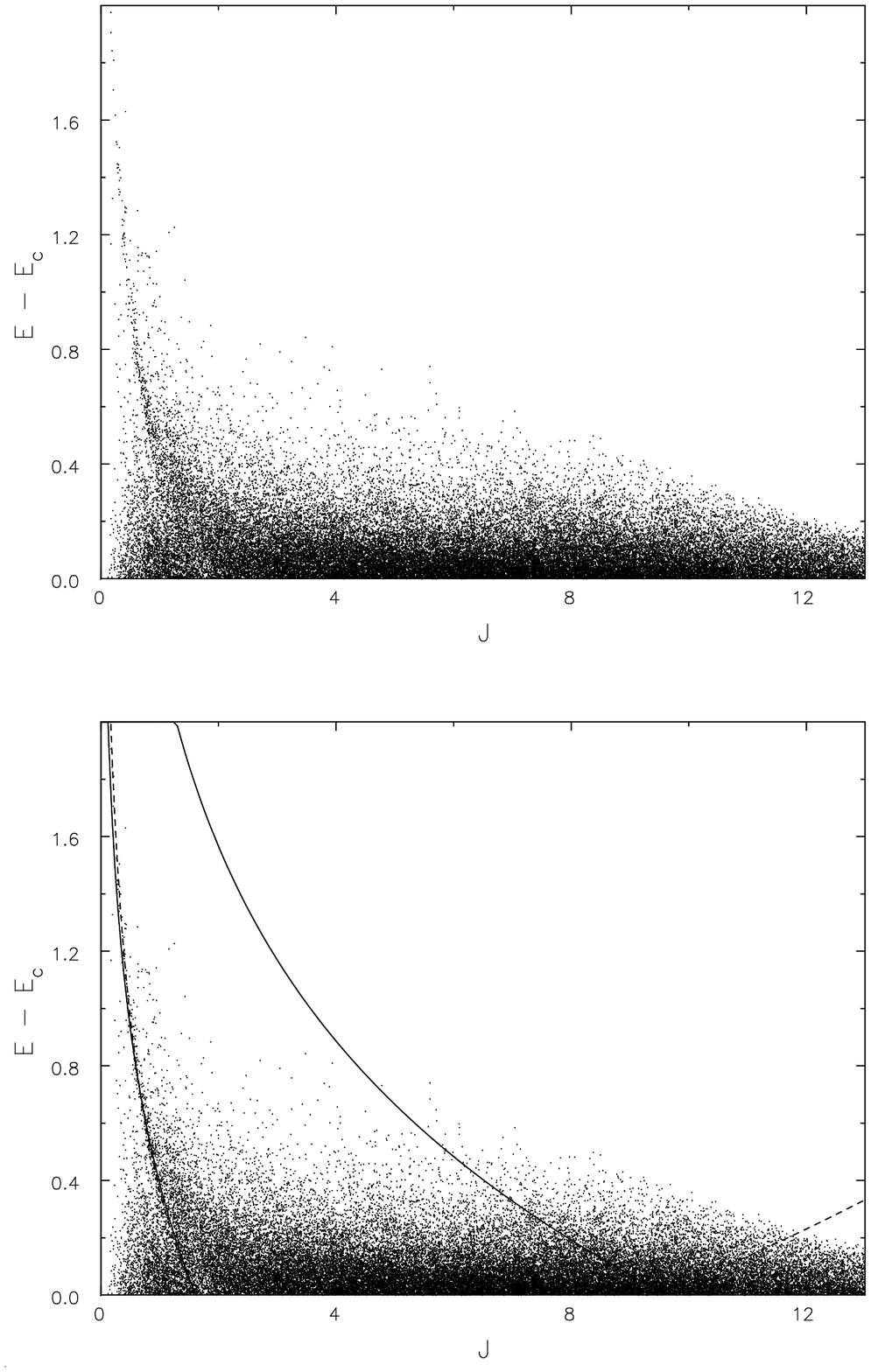}
\caption{The distribution of particles in the space of angular
  momentum and energy of non-circular motion at $t=600$ in the
  simulation mentioned in the text.  The resonant loci (solid) and
  scattering trajectories (dashed) are also shown.}
\label{phasesp}
\end{figure}

Figure~\ref{phasesp} shows the distribution of particles in the space
of angular momentum (here labeled $J$) and energy of random motion
(here labeled $E-E_c$) in this simulation at $t=600$.  It differs from
the initial distribution by the pronounced tongues of particles at low
angular momentum that reach up to high random energies on an upward
curving, negatively sloped trajectory that is almost completely
obscured by the loci of the Lindblad resonances (solid lines), and the
scattering trajectories (dashed lines).  These lines are for the
measured pattern speed of the wave shown in Fig.~\ref{mode} and
therefore have {\it no free parameters}.  The fact that the dashed
line at the \ILR\ overlays the tongue of particles so perfectly that
they are barely visible indicates that these particles were scattered
to more eccentric orbits by the \ILR\ of the identified pattern.
Recall from Fig.~\ref{lindblad} that particles stay on resonance as
they gain angular momentum, which is the reason some reach very high
energies.

\section{Recurrent cycle?}
Each coherent wave leaves behind an altered distribution function and
apparently thereby creates the conditions for a new instability.
Despite having shown this already in \cite{SL89}, I am still unclear
how this happens in detail.  I have not put much effort into pursuing
this question since then, because I was concerned that the behavior I
was studying might turn out to be a horrible artifact of the
simulations and thought it prudent to await evidence that something
similar occurs in nature.  I hoped \citep{Sell94} to see evidence of
resonance scattering in the \Hipp\ measurements of the local stellar
kinematics.

\section{Geneva-Copenhagen Survey}
\cite{Nord04} followed up the accurate distances and proper motions
from \Hipp\ to obtain the missing radial velocities from ground-based
spectra.  Their survey has yielded known distances and full space
motions of over 14\,000 local F \& G dwarfs -- a sample that is
believed to be almost complete within 40 pc and largely free from
selection biases to 100 pc.  I use the latest revision of their survey
\citep{HNA9}, discarding only stars having distances $>200\;$pc and
all known Hyades cluster stars, which leaves a sample of 13\,045
stars.

The addition of radial velocities confirmed the extraordinary
substructure in the $U-V$ plane already discovered by \cite{Dehn98}
in the absence of radial velocities.  The local distribution function
is far from the smooth double Gaussian postulated by Schwarzschild
(BT08, p.~323); there is little evidence for an underlying smooth
component, rather the distribution has the appearance of several
distinct streams \citep{BHR9}, which cannot be dissolved clusters
\citep{Bens07,Fama07,BH09}.  Various groups have attempted to account
for this substructure by resonant scattering by the Milky Way bar
\citep{Dehn00}, spiral arms \citep{DeSi04}, or both
\citep{QM05,Chak07}, which may indeed account for parts of the
observed structure.  But none of these studies was self-consistent,
and all attempted to explain the structure in velocity space.

\begin{figure}[t]
\includegraphics[width=\hsize,clip=]{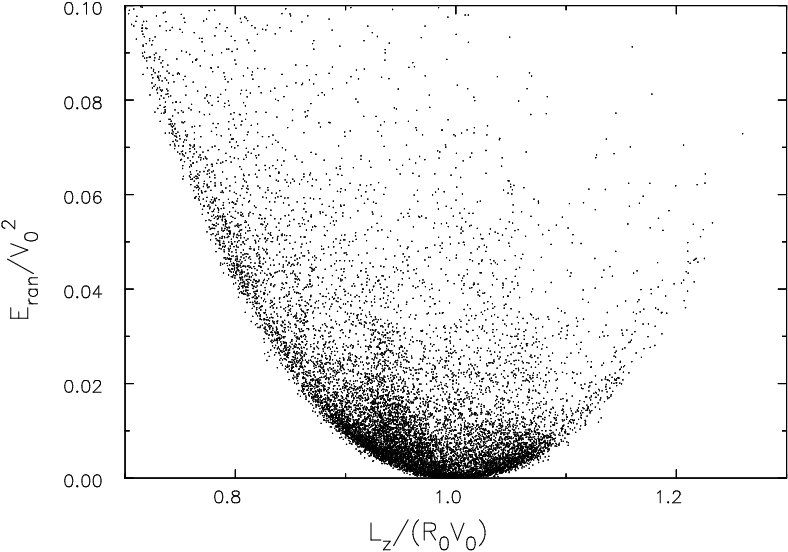}
\caption{The distribution of \GCS\ stars in the space of the two
  integrals $\Eran$ and $L_z$.  Aside from the general decrease in
  density towards large $\Eran$ and a skew to lower $L_z$ caused in
  part by the asymmetric drift, the DF in this projection also shows
  significant substructure. \Ignore{ The solid curve in lower panel
    shows the boundary imposed by selecting stars from the solar
    vicinity only, while the dashed line shows a possible scattering
    trajectory at the \ILR\ for an $m=2$ pattern, and the dot-dash
    line the same for the \OLR.  The dotted line shows the location of
    the \ILR\ for orbits of arbitrary eccentricity and the same
    pattern speed as for the dashed line.}}
\label{intgp}
\end{figure}

It is more instructive to project the data into integral space, as
shown in Figure~\ref{intgp}.  The distribution in the upper panel
assumes a locally flat rotation curve for the Milky Way, $R_0 =
8\;$kpc and $V_0 = 220\;$km/s, but the features in the plot are
insensitive to these assumptions.  The almost parabolic lower boundary
is a selection effect, because stars whose angular momenta differ
increasingly from that of the LSR do not visit the Solar neighborhood
unless they have greater energies of random motion.\footnote{A similar
  excluded region is visible in the quantities plotted by \cite{AF06};
  the omission of the small correction for the different
  Galacto-centric distances of the guiding centers in their
  energy-like term is unimportant.}  The left-right bias results from
the fact that the density of stars in the Galaxy declines outwards,
and therefore larger numbers visit the solar neighborhood from the
innner Galaxy than from larger radii (\ie\ the asymmetric drift).

The distribution in Figure~\ref{intgp} manifests one strong tongue
with a steep negative slope rising from an abscissa $L_z \simeq
0.96R_0V_0$.  This could be the scattering feature at an \ILR\ that I
suggested \citep{Sell94} might be present, but in fact trapping at any
of the principal resonances would cause an excess of stars along a
line of similar slope.  Nevertheless the feature is a highly
significant density excess, with $\gg 99\%$ confidence from a
bootstrap analysis \citep{Sell10}. While there are hints of other
features, none but the obvious one is statistically significant.

\Ignore{The lines in the lower panel of Figure~\ref{intgp} show the
  loci of possible resonances and scattering trajectories for patterns
  having different angular frequencies.  An adjustment to the pattern
  speed slides these lines to the left (higher $\Omega_p$) or right.
  Clearly the inclinations of all the resonance lines, but not the
  scattering line at the \OLR, closely parallel that of the strong
  feature in the upper panel, and therefore one cannot say whether the
  feature is caused by scattering at an \ILR\ or trapping at any
  resonance.}

\begin{figure}[t]
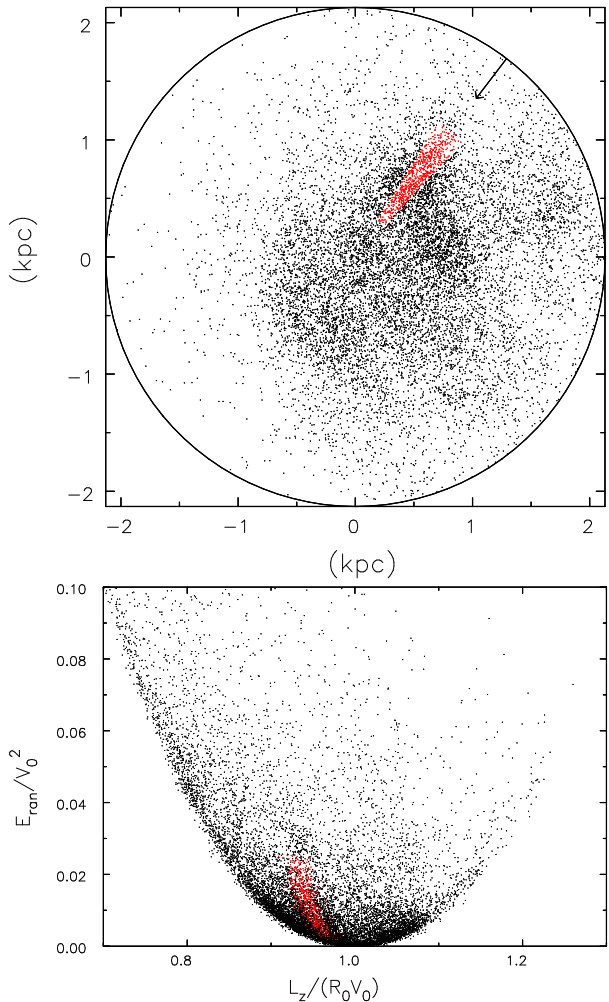

\includegraphics[width=.9\hsize,clip=]{polar.ps}
\includegraphics[width=.9\hsize,clip=]{erancol.ps}
\caption{Top panel shows the distribution of \GCS\ stars in
  action-angle space.  The radial coordinate is the square root of the
  radial action, while the azimuthal coordinate is $2w_\phi - w_R$.
  Note the excess of stars with phase $\sim 53^\circ$ (marked by the
  arrow).  The lower panel shows the distribution in $(L_z,E_{\rm
    ran})$ space.  The red points are the same stars in each plot.}
\label{phases}
\end{figure}

In order to determine which resonance could have caused this feature,
we must examine the phases of the stars, which should exhibit
different correlations depending upon the resonance responsible.  I
therefore computed action-angle coordinates for each star
\citep{Sell10}, and show the results in Figure~\ref{phases}.  The
radial coordinate in the upper panel shows the approximate epicycle
size $a = \sqrt{2J_R/\kappa}$ while the azimuth is $2w_\phi - w_R$.
The concentration of stars at one phase, marked by the arrow is highly
significant.  Furthermore, the stars colored red in the skinny
triangle in the upper panel are exactly the stars, also colored red in
the lower panel, that defined the scattering tongue.  This Figure
therefore confirms that the tongue of stars in integral space has been
caused by scattering at an \ILR.

Unfortunately, it is not possible to tell from this analysis the
angular periodicity of the disturbance that created this feature; it
could be an $m=2$ spiral, but higher angular periodicities would give
rise to very similar features in these plots.  However, an $m=2$ wave
seems the most likely, since the alignment of the scattering
trajectory and resonance locus in Fig.~\ref{lindblad} is closest for
this angular periodicity, making it easier for strong scattering to
produce a clear feature.

The identification of an \ILR\ provides strong support for the picture
of spiral generation that I have been developing from the simulations.
It suggests that spirals in the Milky Way are transient, with the
decay of one pattern seeding a new instability that produces another
spiral pattern.  An \ILR\ also excludes the \cite{BL96} quasi-steady
wave hypothesis as the cause of the pattern that gave rise to this
particular feature, although this one case clearly does not preclude
the possibility that it may still apply elsewhere.

\section{Conclusions}
Globally stable $N$-body simulations have manifested recurrent
transient spiral patterns for many years, and the phenomenon has not
changed as numerical quality has risen.  The transient nature of
spiral patterns receives indirect support both from the importance of
gas and secular heating of disks.  The more recently discovered
further consequence of radial mixing in disks \citep{SB02} is
important for structural evolution, metallicity gradients, dynamo
theory, \etc

Simulations of the only known stable disk with non-uniform rotation
have revealed that linear perturbation theory breaks down at relative
overdensities of just $\sim2\%$.  Real galaxy disks, which contain
both star clusters and molecular clouds, are not that smooth and it
would seem therefore that their behavior cannot be fully described by
linear theory.  I show how non-linear effects could lead to a
recurrent cycle of spirals, although the exact mechanism remains
obscure.

The Geneva-Copenhagen survey has full phase space coordinates of an
unbiased sample of local F \& G dwarf stars.  Examining the
distribution as a function of integrals has yielded clear evidence for
the existence of an inner Lindblad resonance in the solar
neighborhood.  This discovery provides strong support for the
mechanism for recurrent spiral activity that I have been developing.

\section*{Acknowledgments}
I would like to thank Doug Lin once again for the stimulus my research
received during my visit to Santa Cruz twenty-two years ago.  I also
thank Scott Tremaine for helpful e-mail correspondence.  This work was
supported by NSF grant AST-0507323.

\label{lastpage}

\end{document}